\documentclass[aps,prd,twocolumn,showpacs,superscriptaddress,preprintnumbers,nofootinbib,notitlepage]{revtex4-1}
\usepackage[colorlinks=true, pdfstartview=FitV, linkcolor=red,citecolor=blue, urlcolor=blue,
bookmarks=true, bookmarksnumbered=true, breaklinks]{hyperref}
\usepackage[dvipdfmx]{graphicx}
\usepackage{amsmath,amssymb,tabularx,ulem,mathrsfs,bm,multirow}
\begin{document}

\title{Charmonium ground and excited states at finite temperature\\ from complex Borel sum rules}

\author{Ken-Ji Araki}
\email{k.araki@th.phys.titech.ac.jp}
\affiliation{Department of Physics, Tokyo Institute of Technology, Meguro, Tokyo, 152-8551, Japan}
\author{Kei Suzuki}
\email{k.suzuki.2010@th.phys.titech.ac.jp}
\affiliation{Department of Physics, Tokyo Institute of Technology, Meguro, Tokyo, 152-8551, Japan}
\author{Philipp Gubler}
\email{pgubler@riken.jp}
\affiliation{Department of Physics, Keio University, Kanagawa 223-8522, Japan}
\affiliation{Research and Education Center for Natural Science, Keio University, Kanagawa 223-8521, Japan}
\affiliation{Advanced Science Research Center, Japan Atomic Energy Agency, Tokai, Ibaraki, 319-1195, Japan}
\author{Makoto Oka}
\email{oka@th.phys.titech.ac.jp}
\affiliation{Department of Physics, Tokyo Institute of Technology, Meguro, Tokyo, 152-8551, Japan}
\affiliation{Advanced Science Research Center, Japan Atomic Energy Agency, Tokai, Ibaraki, 319-1195, Japan}

\date{\today}

\begin{abstract}
Charmonium spectral functions in vector and pseudoscalar channels at finite temperature are investigated through the complex Borel sum rules and the maximum entropy method.
Our approach enables us to extract the peaks corresponding to the excited charmonia, $\psi^\prime$ and $\eta_c^\prime$, as well as those of the ground states, $J/\psi$ and $\eta_c$, which has never been achieved in usual QCD sum rule analyses.
We show the spectral functions in vacuum and their thermal modification around the critical temperature, which leads to the almost simultaneous melting (or peak disappearance) of the ground and excited states.
\end{abstract}
\maketitle

\section{Introduction}
The particle consisting of a heavy quark and an antiquark $Q\bar{Q}$, ``quarkonium", has been a suitable target to study dynamics of QCD at short distance due to its large mass \cite{Novikov:1977dq}. 
Furthermore, it was suggested and believed that quarkonia in an extremely hot and dense matter, quark gluon plasma (QGP), dissolve due to the color Debye screening caused by light deconfined quarks, so that such an event itself can be regarded as a signal for the existence of QGP \cite{Matsui:1986dk}. Experimentally, this is indeed observed as the quarkonium suppression in heavy ion collisions at Relativistic Heavy Ion Collider (RHIC) at BNL and Large Hadron Collider (LHC) at CERN.
On the theoretical side, this phenomenon can be understood as the thermal modification of $Q\bar{Q}$ potential \cite{Matsui:1986dk,Karsch:1987pv,Karsch:1990wi,Digal:2001iu,Wong:2004zr,Alberico:2005xw,Satz:2005hx,Mocsy:2007yj,Mocsy:2007jz,Laine:2006ns,Rothkopf:2011db} or the disappearance of peaks in the spectral function by the finite temperature effects.
The temperatures at which the peaks completely disappear, ``melting temperatures"~\footnote{Actually, the ``melting temperature" is not a well defined concept because the quarkonium wave function is gradually broadened by thermal effects, and the peak in the spectral function does not necessarily show abrupt disappearance. Therefore, in this work, we use this term just as a qualitative guideline concept.}, are one of the targets which should ideally be calculated from the first principles of QCD. 
Recently, lattice QCD simulations with the maximum entropy method (MEM) \cite{Asakawa:2000tr} enable us to check such a spectral function deformation and to estimate the melting temperature.
For instance, it was shown in such an approach that the lowest charmonium states ($J/\psi$ and $\eta _c$) survive above the temperature of $1.5 \, T_c$ \cite{Asakawa:2003re} (cf.~\cite{Umeda:2002vr,Datta:2003ww,Iida:2006mv,Jakovac:2006sf,Ding:2012sp,Borsanyi:2014vka,Ikeda:2016czj}). 
Similarly, QCD sum rules \cite{Shifman:1978bx,Shifman:1978by} can study quarkonium suppression by incorporating finite temperature effects through the QCD condensates. 
Initially, the ground states of charmonia and bottomonia were analyzed by assuming a specific functional form for the spectral functions \cite{Morita:2007pt,Morita:2007hv,Song:2008bd,Dominguez:2009mk,Morita:2009qk,Dominguez:2010mx,Dominguez:2013fca,Kim:2015xna,Kim:2017pos}, while recently they were reanalyzed without such assumptions by the help of the MEM \cite{Gubler:2011ua,Suzuki:2012ze}.

In the recent heavy-ion collision experiments, the suppression of charmonium excited states ($\psi^\prime$ or $\psi(2S)$) was also observed
\cite{Khachatryan:2014bva,Sirunyan:2016znt,Adam:2015isa,Adam:2016ohd,Adare:2016psx,Aaij:2016eyl}.
The experimental results show that the yield of the excited state is more strongly suppressed when more nucleons participate in the collision. 
This indicates that in QGP the excited state melts at lower temperature than the ground state since the excited state suffers from the Debye screening more strongly due to the larger system size compared to the ground state. 
If this is the case, the second peak in the spectral function should disappear at lower temperature than the first peak also in a theoretical calculation.
Actually some model studies demonstrate such a situation \cite{Matsui:1986dk,Karsch:1987pv,Karsch:1990wi,Digal:2001iu,Wong:2004zr,Alberico:2005xw,Satz:2005hx,Mocsy:2007yj,Mocsy:2007jz}. 
Also the studies of bottomonia at finite temperature using QCD sum rules with MEM find an indirect proof indicating the same effect, without explicitly reproducing the peaks corresponding to the excited sates \cite{Suzuki:2012ze}.

In general, it is a challenging problem to extract information on excited states (namely, second or higher peaks in the spectral function) from QCD sum rules.
Therefore, the thermal modification of excited states have so far not been discussed within the {\it usual} Borel type QCD sum rules, although there exist many results for the ground states.
Although MEM for QCD sum rules is a powerful tool to investigate the structure of spectral function, which has already been applied to various systems \cite{Gubler:2010cf,Gubler:2011ua,Ohtani:2011zz,Suzuki:2012ze,Ohtani:2012ps,Gubler:2014pta,Suzuki:2015est,Ohtani:2016pyk}, it was difficult to reproduce the second peak with statistical significance from Borel sum rules.
However, by using the combination of QCD sum rules on the complex Borel plane (CBSR) and MEM, it has recently become possible to extract excited states due to the improved resolution \cite{Araki:2014qya}.
In Ref.~\cite{Araki:2014qya}, the excited $\phi$ meson peak in vacuum was reproduced with a mass consistent with the experimental value.
The reproduction of excited charmonia {\it from QCD sum rules} is also one of main purposes of this paper.

In this work, we study charmonia in the vector and pseudoscalar channels at finite temperature by using CBSR with MEM, which has the ability to reproduce both the first and second peaks.
Then we check whether there is a possible difference in the melting behaviors between the ground and excited states.

This paper is organized as follows.
In Section \ref{Sec_Formalism}, we explain the CBSR for charmonia at finite temperature.
In Section \ref{Sec_Results}, we show their spectral functions in vacuum and at finite temperature and discuss their thermal (melting) behavior.
Section \ref{Sec_Conclusion} is devoted to our conclusion and outlook.

\section{Formalism} \label{Sec_Formalism}
Let us here introduce our formalism of CBSR with MEM for the analysis of charmonium at finite temperature.
Since CBSR is a generalization of real Borel sum rules even at finite temperature, we can take the same Borel sum rule formulation as done in the previous works, where we replace real Borel masses by complex ones, to construct CBSR.
Here we will briefly summarize our formulation; a more detailed explanation can be found in Refs.~\cite{Gubler:2011ua,Suzuki:2012ze}.
For specifying the domain of the complex Borel plane to be used in the MEM analysis, we propose an updated criterion. 
 
\subsection{The form of Borel sum rules}
We calculate the correlation function for a meson system consisting of charm quarks with a mass $m$ that is larger than the typical QCD scale.
Here, the dimensionless correlation functions in momentum space are defined by $\tilde{\Pi}^V(q^2) \equiv \Pi_\mu^{V, \mu}(q) /(-3q^2)$ and $\tilde{\Pi}^P \equiv \Pi^P (q)/q^2$ for vector and pseudoscalar channels, respectively.
These can be calculated by the operator product expansion (OPE), and its Borel-transformed OPE with a real variable $M^2$ takes the following shortened form: 
\begin{eqnarray}
G^J(M^2; T) &=&  \frac{1}{M^2} \mathrm{e}^{-\nu} A^J(\nu) \nonumber\\
&& \hspace{-40pt} \Big[ 1+  \alpha_s(\nu) a^J(\nu)  +b^J(\nu) \phi _b (T) +c^J(\nu) \phi_c(T) \Big],
\label{eq:Tope}
\end{eqnarray}
where $\nu$ is a dimensionless parameter defined as $\nu =4m^2/M^2$. The superscript $J$ distinguishes the channel, vector or pseudoscalar.
With this expression we construct sum rules at finite temperature as follows:
\begin{equation}
G^J(M^2; T)= \frac{1}{M^2}\int ^{\infty}_0 \mathrm{e}^{-s/M^2}\rho(s;T)ds.
\label{eq:Tdis}
\end{equation}

Let us explain each part of Eq.~(\ref{eq:Tope}).
$\alpha_s (\nu)$ is a strong running coupling constant evaluated at the Borel mass scale $M$. The functions, $A^J, \ a^J,\ b^J\ \mbox{and} \ c^J$ are the Wilson coefficients. Their explicit form can be found in Ref.~\cite{Morita:2009qk}. The first and second terms come from the usual perturbative result up to the first order in $\alpha_s$.
The $\phi (T)'s$ are the condensates defined as
\begin{eqnarray}
\phi _b (T) &=& \frac{4 \pi ^2}{9 (4 m^2)^2}G_0 (T),\\
\phi _c (T) &=& \frac{4 \pi ^2}{3 (4 m^2)^2}G_2 (T),
\end{eqnarray}
where $G_0(T)$ and $G_2(T)$ are the scalar and twist-2 gluon condensates with dimension 4 at finite temperature.
They are defined as the scalar and spin-2 components of the gauge independent gluon tensor:
\begin{eqnarray}
  \langle \frac{\alpha_s}{\pi} G^{a\mu \sigma} G^{a \  \nu}_{\ \sigma} \rangle_T&=& \frac{1}{4}g^{\mu \nu} G_0(T) + (u^{\mu}u^{\nu}  -\frac{1}{4} g^{\mu\nu}) G_2(T), \nonumber \\ 
\end{eqnarray}
where the expression $ \langle O \rangle_T$ means the Boltzmann average defined as $ \langle O \rangle_T= \mathrm{Tr}(\mathrm{e}^{-H/T}O  )/\mathrm{Tr}(\mathrm{e}^{-H/T} )$.
$u$ is the four velocity vector of the medium whose norm is unity. 

\subsection{Finite temperature effects}
In this work we assume that all temperature dependence of the correlator enters through that of the gluon condensates.
This assumption is valid as long as the temperature is smaller than the OPE separation scale, which is of the order of the Borel mass $M \sim 1, \mathrm{GeV}$.
To proceed further, we therefore need to calculate the temperature dependences of the gluon condensates.
In our formulation, the strategy proposed in Refs.~\cite{Morita:2007pt,Morita:2007hv} is employed, in which we express the two gluon condensates as follows:
\begin{eqnarray}
G_0(T)&=& G_0(0) -\frac{8}{11}[ \epsilon(T) -3p(T)], \label{eq:G0.temp} \\
G_2(T)&=& -\frac{\alpha_s(T)}{\pi} [\epsilon(T) +p(T)], \label{eq:G2.temp}
\end{eqnarray}
where $\epsilon$, $p$ and $\alpha_s$ are the energy density, pressure and strong coupling constant, respectively.
Thus by estimating their dependences on temperature by using quenched (pure Yang-Mills) lattice QCD simulations \cite{Boyd:1996bx,Kaczmarek:2004gv}, we finally get the temperature dependences of the condensates.
Actually the value of the critical temperature of the chiral phase transition in the quenched approximation is about $260 \, \mathrm{MeV}$, while the full QCD result leads to the cross-over temperature being $145-165 \, \mathrm{MeV}$ \cite{Borsanyi:2010bp,Bazavov:2011nk}.
Such a quantitative difference caused by the quenched approximation should be discussed in future works.
 
Let us here briefly discuss the possibility of extending our approach to full QCD, which would include dynamical quarks. 
Eqs.\,(\ref{eq:G0.temp}) and (\ref{eq:G2.temp}) in fact are derived by matching the trace part and the symmetric traceless part of the energy momentum 
tensor, written either in terms of thermodynamic variables or the basic degrees of freedom of QCD. While in the quenched approximation, the energy momentum 
tensor of QCD can be written only with gluon fields, it will have terms such as $m_q \langle \bar{q} q \rangle$ and 
$\langle \bar{q} \gamma^{\mu} D^{\nu} q \rangle$, which will hence modify Eqs.\,(\ref{eq:G0.temp}) and (\ref{eq:G2.temp}). Once the 
temperature dependences of all these operators (and $\alpha_s$) are known in full QCD, it will become possible to perform 
a QCD sum rule calculation that goes beyond the quenched approximation.

\subsection{The choice of the domain in the complex Borel mass plane }
CBSR can be constructed simply by replacing the Borel mass $M^2$ in Eqs.~(\ref{eq:Tope}) and (\ref{eq:Tdis}) by its complex generalization $\mathcal{M}^2$.
As a further task, we have to choose the domain of complex Borel masses used in the MEM analyses.
Because spectral functions obtained from MEM generally depend on the choice of the domain, the domain used for each channel in our analyses should be kept constant throughout the analyses at various temperatures, such that the temperature dependence of the spectral function can be extracted without artificial effects.
In our analyses, the domain is fixed to the one used in vacuum for each channel.
By considering the convergence of the OPE, the lower boundaries are determined as follows:  
\begin{equation}
\frac{|d^4(\mathcal{M}^2;T=0)|}{|G^J( \mathcal{M}^2;T=0)|}\ < \ 0.1,
\label{eq:inner}
\end{equation}
where $d^4(\mathcal{M}^2;T=0)$ is the whole dimension 4 term of the OPE in vacuum~\footnote{
This method of setting the lower boundaries was proposed in the first analysis using CBSR with MEM \cite{Araki:2014qya}.}. 

\begin{figure}[t!]
    \centering
    \includegraphics[clip, width=1.0\columnwidth]{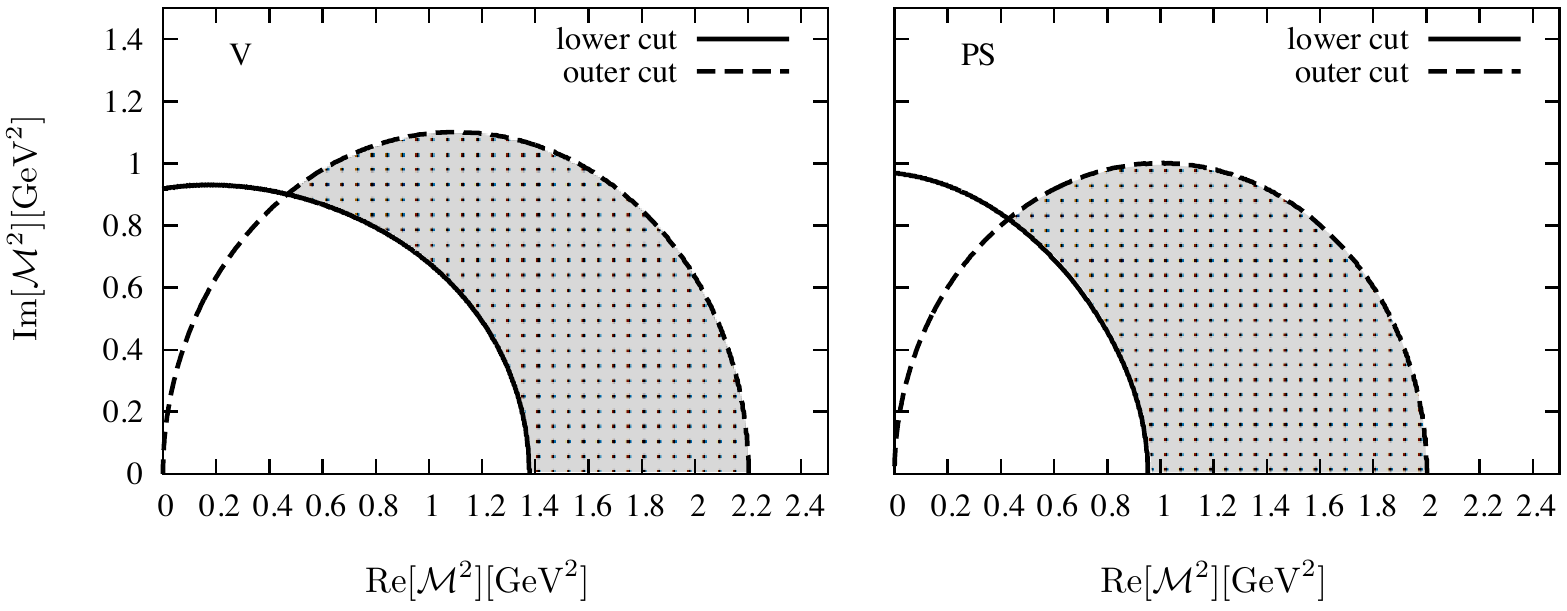}
    \caption{Domains in the complex Borel mass plane used for the MEM analysis in both vector (left plot) and pseudoscalar (right plot) channels. The lower and upper boundaries are determined according to Eq.~(\ref{eq:inner}) and Eq.~(\ref{eq:outer}), respectively.}
    \label{domain}
\end{figure}

For the determination of the upper boundary, we employ an improved scheme as follows.
In the originally proposed scheme \cite{Araki:2014qya}, we employed a simple circular boundary on the complex Borel mass plane, whose radius $M_r^2$ is treated as a free parameter as follows: $M^2 < M^2_r$, where $\mathcal{M}^2= M^2 \mathrm{e}^{i \theta}$.
However, it includes the region $\theta \sim \frac{\pi}{2}$, where the damping by the kernels becomes weaker, as shown in the definitions
\begin{equation}
\begin{split}
K^{\mathrm{R}}(\mathcal{M}^2;s) & \equiv \mathrm{Re}\Bigl[ \, \frac{1}{\mathcal{M}^2} \, \mathrm{e}^{-s/\mathcal{M}^2} \, \Bigr]\\
& =\frac{1}{M^2}  \mathrm{e}^{-( \cos{\theta}/M^2)s }  \cos{\bigl[ \, (\sin{\theta} /M^2)s - \theta \, \bigr]},
\label{eq:reker}
\end{split}
\end{equation}
\begin{equation}
\begin{split}
K^{\mathrm{I}}(\mathcal{M}^2;s) &  \equiv \mathrm{Im}\Bigl[  \, \frac{1}{\mathcal{M}^2} \, \mathrm{e}^{-s/\mathcal{M}^2} \, \Bigr]\\
&=\frac{1}{M^2}  \mathrm{e}^{-( \cos{\theta}/M^2)s }  \sin{\bigl[ \, (\sin{\theta} /M^2)s - \theta \, \bigr]}.
\label{eq:imker}
\end{split}
\end{equation}
Because, in such a region, the integrals over the spectral function include large contributions from the continuum, it is natural to set a lower limit for the power of the exponential in these kernels as follows:
\begin{equation}
\frac{\cos \theta}{M^2} \ >\  \tilde{r}_{c}.
\label{eq:outer}
\end{equation}
Such a limit can adequately control the contribution from higher energy regions of the spectral function.
It leads to a curve boundary on the complex Borel mass plane, as shown by the dashed line in Fig.~\ref{domain}. 
We choose the critical value $\tilde{r}_{c}$ such that the statistical significances of both the first and second peaks in vacuum become best.
As a result, $\tilde{r}_c$ is equal to $1/2.2 \, \mathrm{GeV}^{-2}$ and $1/2.0 \, \mathrm{GeV}^{-2}$ for the V and PS channels, respectively.
The actual domains for the V and PS channels determined by the above conditions, Eq.~(\ref{eq:inner}) and Eq.~(\ref{eq:outer}), are shown in Fig.~\ref{domain} as shaded regions, in which the discretized complex Borel masses are used as input for our MEM analyses.  

The default model, which is another MEM input (see e.g. Ref.~\cite{Asakawa:2000tr}), is chosen as a constant 
fixed to the asymptotic value of the spectral function computed by perturbation theory \cite{Morita:2009qk}: the value of $\frac{1}{\pi} \mathrm{Im} \Pi_{pert.}^{V,PS} (\omega^2)$ at $\omega=\infty$ for the V channel and $\omega=10 \, \mathrm{GeV}$ for the PS channel~\footnote{Because $\mathrm{Im}\Pi_{pert.}^{PS} (\infty)$ diverges, we are forced to choose a finite energy to determine the ``asymptotic value" of the spectral function. We checked that the obtained peaks do not strongly depend on the choice of this parameter.}.
The values of the other input parameters used in our analyses are summarized in Table~\ref{table:condensate}.

\begin{table}[t!]
\begin{center}
\begin{tabular*}{1.0\columnwidth}{@{\extracolsep{\fill}}| c||  c  c|}
\hline
$\bar{m}_c(\bar{m}_c) $                                 &  $ 1.273 \pm 0.006 $  \cite{McNeile:2010ji}   &        \\
\hline
 $\ \ \   \langle\frac{ \alpha_s}{\pi}G ^2\rangle  \ \ \ $        &  $0.012 \pm 0.0036 \ \mathrm{GeV}^4$  \cite{Shifman:1978by,Colangelo:2000dp}      &     \\
\hline
 $   \Lambda_{QCD} $                         &  $   0.213 \pm 0.008 $   \cite{Bethke:2012jm}         &     \\
\hline
\end{tabular*}
\end{center}
\caption{Values and respective uncertainties of the condensates and other parameters used for evaluating the 
OPE of Eq.~(\ref{eq:Tope}).}
 \label{table:condensate}
\end{table}

\section{Analysis results} \label{Sec_Results}
\subsection{In vacuum}
\begin{figure}[t!]
 \begin{center}
  \includegraphics[clip, width=1.0\columnwidth]{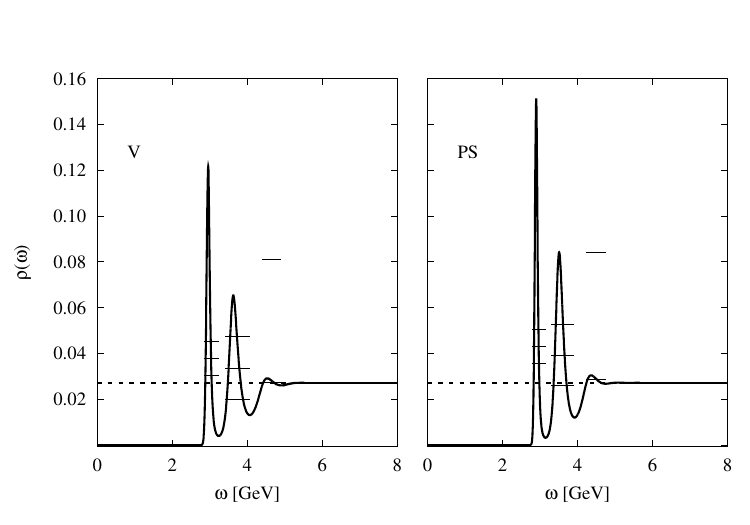}
 \end{center}
 \caption{The analysis results of the vector (left plot) and pseudoscalar (right plot) channels in vacuum.
The solid lines show the spectral function extracted from MEM. The dashed lines show the default model.}
 \label{cc_vac}
\end{figure}

The spectral functions of both V and PS channels in vacuum obtained by our analyses are shown in Fig.~\ref{cc_vac}.
The estimated errors of the MEM results are shown by the three horizontal lines at each peak.
It is observed that in both channels two clear peaks are generated.
Both of them are statistically significant because their error bars lie between top and bottom of the peaks \cite{Asakawa:2000tr,Gubler:2010cf}.
Note that, although a small third peak is also seen in both channels, it could be an artificial peak because it is not statistically significant.
The positions of the first and second peaks agree with the experimental values with a precision of the order $50-150 \, \mathrm{MeV}$ as shown in Table~\ref{position}.
Thus they are considered to correspond to the physical states, $J/\psi$, $\psi^\prime$, $\eta_c$ and $\eta_c^\prime$, respectively. 
It is interesting to note that all our obtained masses are lower compared to the experimental values. It is possible that this situation 
could improve once higher order $\alpha_s$ corrections to the Wilson coefficients are included in the analysis. As such terms have no influence on the temperature dependence of the gluon condensates, it is however not expected that they will qualitatively change our results about the 
temperature dependence of the peaks.

Let us note that, in an earlier study using conventional Borel sum rules with MEM \cite{Gubler:2011ua}, only one statistically significant peak was extracted for both channels, and it is widely distributed in the energy region between about $2.9 \, \mathrm{GeV}$ and $3.6 \, \mathrm{GeV}$. 
This wide peak likely corresponds to a combination of the first and second peaks obtained in our analysis. 
In other words, by the help of the higher resolution of CBSR, we are able to separate the degenerated of the previous work into two distinct peaks.
%

\begin{table}[t!]
\begin{center}
\begin{tabular*}{1.0\columnwidth}{@{\extracolsep{\fill}}c cccc}
 \hline
 \multirow{2}{*}{}        &\multicolumn{2}{c}{ V }& \multicolumn{2}{c}{PS}\\
\hline
&1st [GeV] &2nd [GeV] &1st [GeV]  &2nd [GeV] \\ 
 \hline
 \hline
This work  & 2.954 & 3.624 & 2.900 & 3.512 \\
Experiment & 3.096 & 3.686 & 2.983 & 3.639 \\
 \hline
\end{tabular*}
 \caption{Positions of the charmonia peaks in vacuum, extracted from CBSR and MEM. The corresponding spectral functions are shown in Fig.~\ref{cc_vac}.}
 \label{position}
  \end{center}
\end{table}

\begin{figure}[tbh!]
 \begin{center}
  \includegraphics[clip, width=1.0\columnwidth]{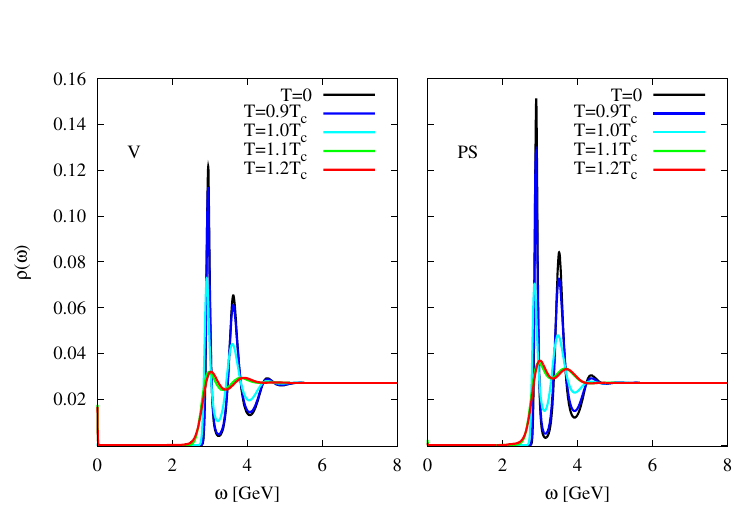}
  \includegraphics[clip, width=1.0\columnwidth]{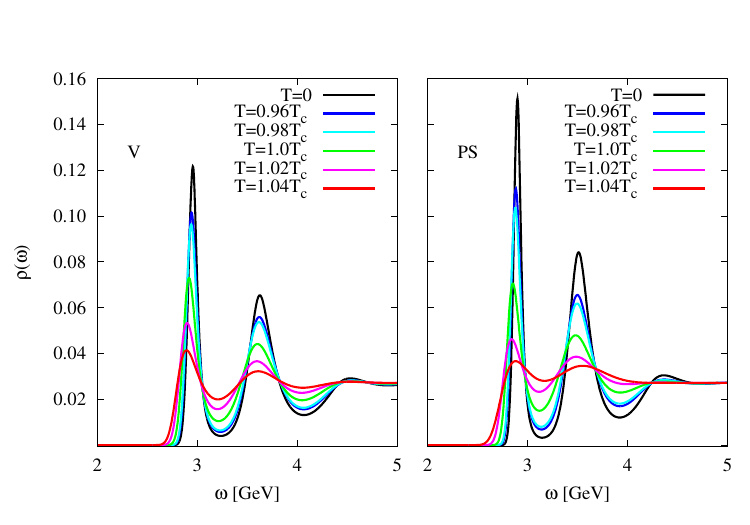}
 \end{center}
  \caption{The temperature dependence of the spectral functions in vector (left plot) and pseudoscalar (right plot) channels.
The temperature ranges between $0.9 \, T_c$ and $1.2 \, T_c$ (top) or $0.96 \, T_c$ and $1.04 \, T_c$ (bottom).
In the bottom figure, the plotted range of horizontal axis is limited to $\omega = 2 \, \mathrm{GeV}$ to $5 \, \mathrm{GeV}$.}
 \label{temp1}
\end{figure}

\subsection{At finite temperature}
Let us see how the peaks are deformed by finite temperature effects.
Fig.~\ref{temp1} shows the spectral functions in the temperature range between $0.9 \, T_c$ and $1.2 \, T_c$ and also in vacuum for comparison.
We observe in the figure that all the peaks tend to disappear with increasing temperature, that is, the charmonia melt by the temperature effects. In both channels, the spectral functions seem to be almost unchanged in the temperature range between $1.1 \, T_c$ and $1.2 \, T_c$.
This observation may be interpreted as the fact that the complete melting of both states up to $1.1 \, T_c$.
On the other hand, the drastic changes of the spectral functions occur around at $1.0 \, T_c$ for both peaks.
This is understood to be caused by a sudden change of the $G_0(T)$ and $G_2(T)$ at about $1.0 T_c$ (see e.g. Ref.~\cite{Morita:2007pt}).

Since we are most interested in comparing the melting behaviors of ground and excited states, we next examine them more carefully in the temperature region around $1.0 \, T_c$.
The bottom panels of Fig.~\ref{temp1} hence show spectral functions at temperatures between $0.96 \, T_c$ and $1.04 \, T_c$ with a interval of $0.02 \, T_c$, while the energy range of the plot is limited to $\omega = 2 - 5 \, \mathrm{GeV}$. 
In this figure, however, we do not see any considerable difference between the ground and excited peaks: Both peaks seem to melt almost similarly with increasing temperature.
On the other hand, in view of the MEM error estimation, we find that the second peak loses its statistical significance at a slightly lower temperature than that of the first peak, as shown in Table~\ref{lower limit}.

\begin{figure}[t!]
    \centering
    \includegraphics[clip, width=1.0\columnwidth]{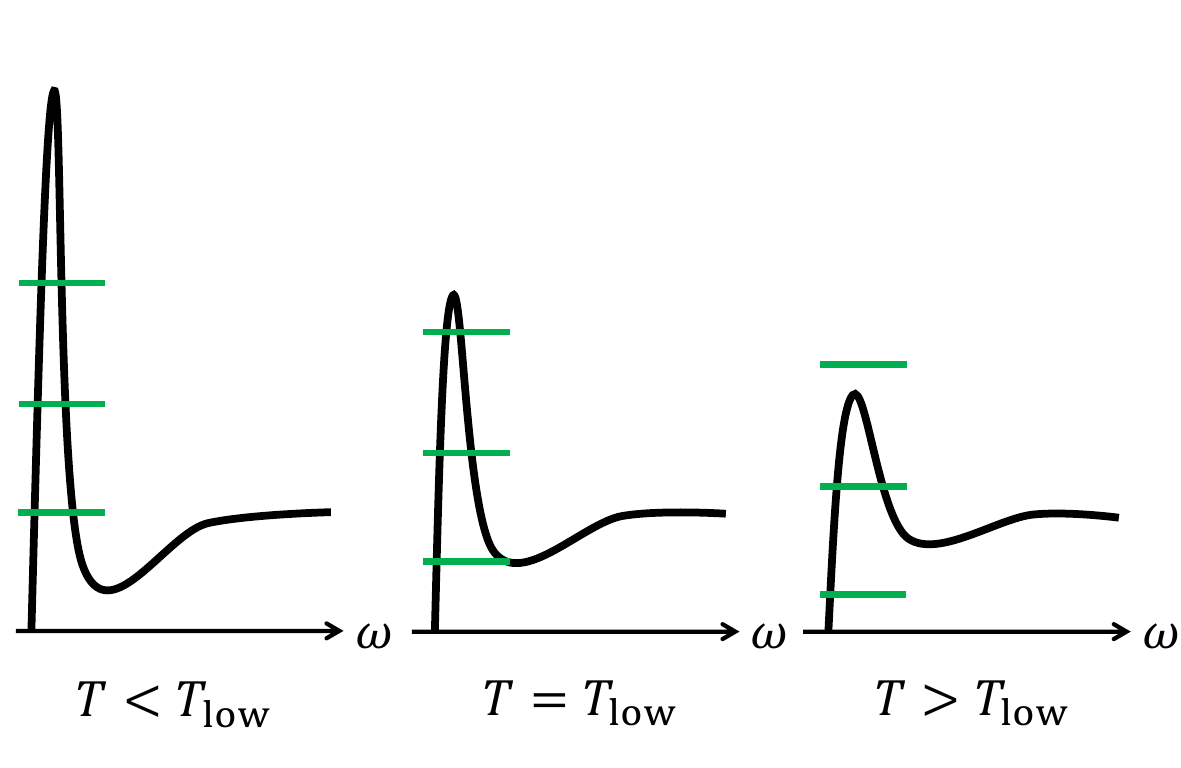}
    \caption{
Typical behaviors of spectral function error bars at finite temperature, extracted from the MEM analysis.
Left ($T< T_\mathrm{low}$): Statistically significant peak.
Middle ($T = T_\mathrm{low}$): Our definition of the lower limit of the melting temperatures, given in Table~\ref{lower limit}.
Right ($T > T_\mathrm{low}$): Statistically non-significant peak.}
    \label{error}
\end{figure}

We note that the temperatures in Table~\ref{lower limit} do not necessarily correspond to the original melting temperatures, but only indicate their ``lower limits".
In Fig.~\ref{error}, we show typical behaviors of error bars at finite temperature.
At lower temperature $T<T_\mathrm{low}$, the peak has statistical significance, so that we can clearly conclude that it survives.
At $T=T_\mathrm{low}$, the lower error bar of the peak reaches ``the dip" at the high-energy side of the peak.
At higher temperature $T>T_\mathrm{low}$, the statistical significance of the peak is lost, so that we cannot conclude anything about its existence.
Thus, all we can say is that the physical peak survives at least up to $T=T_\mathrm{low}$, and $T_\mathrm{low}$ merely indicates a lower limit of the original melting temperature.
Therefore, the original melting temperatures might become larger than the lower limits in Table~\ref{lower limit}.

As shown in Fig.~\ref{temp1} it is worthwhile to note that the positions of both peaks shift to the lower energy side of the order of $50 \, \mathrm{MeV}$ before they lose their statistical significances.
Such a shift was already obtained in earlier studies using Borel sum rules with/without MEM only for the ground states \cite{Morita:2007pt,Morita:2007hv,Song:2008bd,Dominguez:2009mk,Morita:2009qk,Dominguez:2010mx,Dominguez:2013fca,Kim:2015xna,Kim:2017pos,Gubler:2011ua,Suzuki:2012ze} \footnote{We note that the scattering term used in Ref.~\cite{Dominguez:2009mk}, as shown in Appendix A of the reference, has an error.}.
In our analyses by CBSR with MEM, the same behavior is observed also for the excited states.

\begin{table}[t!]
\begin{center}
  \begin{tabular*}{1.0\columnwidth}{@{\extracolsep{\fill}}cccc}
  \hline
    \multicolumn{2}{c}{ V }& \multicolumn{2}{c}{PS}\\
    \hline  
  1st peak\ [$T_c$] &2nd peak\ [$T_c$] &1st peak\ [$T_c$]  &2nd peak\ [$T_c$] \\ 
    \hline
    \hline
  1.01 & 0.99&  1.01& 0.99\\
    \hline
  \end{tabular*}
  \caption{Lower limits of the melting temperature determined by the MEM error estimation.}
  \label{lower limit}
\end{center}
\end{table}

In MEM analyses, the shape of the obtained spectral function generally depends on the error of the input parameters.
It is known that if the error grows larger, the spectral function tends to approach the default model, resulting in an unphysical suppression of a potential peak.
In our QCD sum rule analysis with MEM, the error of the whole OPE becomes larger with increasing temperature through the growing uncertainties of $G_0(T)$ and $G_2(T)$.
To test how large such unphysical effects are, we repeated the same analyses at finite temperature by keeping the error fixed to $T=0$~\footnote{A similar analysis was performed also in Ref.~\cite{Gubler:2011ua}.}.
By comparing such a test analysis and the original one at each $T$, we confirmed that the behavior of the respective spectral functions is not changed.
Therefore, we can conclude that the deformation of spectral function obtained by our analyses is caused not by unphysical effects from the errors but by physically meaningful ones caused by the changing condensate values.

As a more quantitative analysis, we could fit the obtained spectral functions by a functional form with some fitting parameters (e.g. two peaks and continuum), as performed in Ref.~\cite{Suzuki:2012ze}.
However we did not obtain a stable solution for the temperature dependence of residues due to the appearance of local minima around $T_c$, where the spectral structures are more complex than the form of one peak and continuum as the previous study \cite{Suzuki:2012ze}.
Thus we do not discuss residues at finite temperature in this paper.

\section{Conclusion and outlook} \label{Sec_Conclusion}
We have applied CBSR with MEM to the vector and pseudoscalar channels of charmonium spectra in vacuum and at finite temperature.
With the help of the higher resolution of CBSR, two statistically significant peaks in both channels are extracted, which were not resolved in the earlier Borel sum rules with MEM analysis of Ref.~\cite{Gubler:2011ua}.
Their peak positions agree well with the experimental values of $J/\psi$, $\psi^\prime$, $\eta_c$ and $\eta_c^\prime$ within $ 50 - 150 \, \mathrm{MeV}$. 
By introducing finite temperature effects through two dimension-4 gluon condensates, we have observed that all the obtained peaks are deformed to gradually disappear as the temperature increases.
We have confirmed that this is not an artificial MEM effect induced by the increasing error.
They completely melt at the temperature of $T=1.1 \, T_c$ since the spectral function does not largely change between $1.1 \, T_c$ and $1.2 \, T_c$ compared with drastic change around at $1.0 \, T_c$.
In both the channels, the first and second peaks seem to melt almost simultaneously. 
It is interesting to compare our predictions for excited states with recent experimental \cite{Khachatryan:2014bva,Sirunyan:2016znt,Adam:2015isa,Adam:2016ohd,Adare:2016psx,Aaij:2016eyl} or theoretical results \cite{Aarts:2011sm,Burnier:2015tda,Braga:2016wkm}. 

To further improve the QCD sum rules, we can include higher dimensional terms in the OPE and their temperature dependences.
For example, recently, the temperature dependences of the dimension-6 gluon condensates were phenomenologically estimated in Refs.~\cite{Kim:2015xna,Kim:2017pos}.
In the future, it would be worthwhile to precisely determine such higher dimensional condensates from lattice QCD simulations and to analyze QCD sum rules including these effects.

Finally, we comment on the possibility of studying bottomonium channels within the same approach. 
Recent experimental results from LHC show that the excited states, $\Upsilon(2S)$ and $\Upsilon(3S)$, are more strongly suppressed than the ground state $\Upsilon(1S)$ \cite{Chatrchyan:2011pe,Chatrchyan:2012lxa,Chatrchyan:2013nza,Khachatryan:2016xxp}.
Although we have tried to reproduce the bottomonium spectra using CBSR with MEM, we just obtained one single peak, generated from the overlap of the three original 
peaks, meaning that the resolution of MEM was not good enough to disentangle the three states for the bottomonium case.
This is the same situation as usual Borel sum rules with MEM in our previous study \cite{Suzuki:2012ze}.
The reason for this difficulty is related to the fact that the ratio of the mass differences between $\Upsilon(1S)$, $(2S)$, and $(3S)$ to the typical bottomonium mass scale $\sim 10 \mathrm{GeV}$ is too small to separate these states by our approach, which is different from the analyses of charmonia with the mass difference of $m_{2S} - m_{1S} > 500$ MeV and the energy scale $\sim 3 \mathrm{GeV}$.
The task of further improving the resolution of CBSR and finally separating the bottomonium states is left for future studies.

\begin{acknowledgments}
We are grateful to Kenji Morita for providing us the gluon condensate data extracted from lattice QCD simulations.
K.J.A. was supported by Grant-in-Aid for JSPS Fellows from Japan Society for the Promotion of Science (JSPS) (No. 15J11897).
The research of P.G. is supported by the Mext-Supported Program for the Strategic Foundation at Private Universities, ``Topological Science" (No. S1511006).
This work is supported by Grants-in-Aid for Scientific Research from JSPS [Grant No. JP25247036(A)].
\end{acknowledgments}

\bibliography{ref_sum}

\end{document}